\documentclass[pre,twocolumn,nofootinbib]{revtex4-1}
\usepackage{inconsolata}
\usepackage[bbgreekl]{mathbbol} 
\usepackage{amsfonts,amsthm,amsmath,amssymb,mathrsfs,mathtools}
\usepackage{dsfont}
\usepackage{epsfig,color} 
\usepackage{verbatim,subfigure,framed,graphics,graphicx,wrapfig,textcomp}
\usepackage{epstopdf}
\usepackage{xr}

\newcommand{\be}{\begin{eqnarray}}
\newcommand{\ee}{\end{eqnarray}}
\newcommand{\D}{\mathrm{d}}

\newcommand{\f}{f}

\newcommand{\pphi}{\phi}
\newcommand{\uu}{g}

\newcommand{\dt}{\Delta t}
\newcommand{\xibar}[1]{\bar{\xi}(#1)}
\newcommand{\mom}{k}
\newcommand{\lagr}{\mathcal{L}}
\newcommand{\ham}{\mathcal{H}}
\newcommand{\Smin}{S_{\rm min}}
\newcommand{\Sg}{S_{\rm G}}
\newcommand{\kn}{\tilde{\mom}}

\newcommand{\Vc}{V'_{\rm th}}
\newcommand{\lambdatildec}{\lambdatilde_{\rm th}}
\newcommand{\lambdatilde}{\lambda}
\newcommand{\atilde}{a}
\newcommand{\lambdaorig}{\lambda_0}
\newcommand{\aorig}{a_0}
\newcommand{\SI}{Appendix}

\renewcommand{\eqref}[1]{({\ref{#1}})}
\newcommand{\eqqref}[2]{({\ref{#1}},{\ref{#2}})}
\newcommand{\eqqqref}[3]{({{\ref{#1}},{\ref{#2}},{\ref{#3}}})}
\newcommand{\eqtoref}[2]{({\ref{#1}}--{\ref{#2}})}

\begin{document}

\title{Exponential increase of transition rates in metastable systems driven by non-Gaussian noise}

% Use letters for affiliations, numbers to show equal authorship (if applicable) and to indicate the corresponding author
\author{Adrian Baule\textsuperscript{1} and Peter Sollich\textsuperscript{2,3}} 

\affiliation{\textsuperscript{1}School of Mathematical Sciences, Queen Mary University of London, London E1 4NS, UK\\
\textsuperscript{2}Institute for Theoretical Physics, Georg-August-University G\"ottingen, D-37077 G\"ottingen, Germany\\
\textsuperscript{3}Department of Mathematics, King's College London, London WC2R 2LS, UK}

\begin{abstract}
Non-Gaussian noise influences many complex out-of-equilibrium systems on a wide range of scales such as quantum devices, active and living matter, and financial markets. Despite the ubiquitous nature of non-Gaussian noise, its effect on activated transitions between metastable states has so far not been understood in generality, notwithstanding prior work focusing on specific noise types and scaling regimes. Here, we present a unified framework for a general class of non-Gaussian noise, which we take as any finite-intensity noise with independent and stationary increments. Our framework identifies optimal escape paths as minima of a stochastic action, which enables us to derive analytical results for the dominant scaling of the escape rates in the weak-noise regime generalizing the conventional Arrhenius law. We show that non-Gaussian noise always induces a more efficient escape, by reducing the effective potential barrier compared to the Gaussian case with the same noise intensity. Surprisingly, for a broad class of amplitude distributions even noise of infinitesimally small intensity can induce an exponentially larger escape rate. As the underlying reason we identify the appearance of discontinuous minimal action paths, for which escape from the metastable state involves a finite jump. We confirm the existence of such paths by calculating the prefactor of the escape rate, as well as by numerical simulations. Our results highlight fundamental differences in the escape behaviour of systems subject to thermal and non-thermal fluctuations, which can be tuned to optimize switching behaviour in metastable systems.
\end{abstract}

\maketitle

Activated transitions between metastable states govern a large variety of phenomena in the physical, chemical, and biological sciences, ranging from chemical reactions to nucleation, self-assembly, and protein folding \cite{Hanggi:1990aa,Bryngelson:1987aa,Auer:2001aa,Whitelam:2015aa}. Following seminal works by Arrhenius, Eyring and Kramers, the description of transition rates has been well understood for systems at thermal equilibrium, for which the noise driving the transition is Gaussian: transition rates can be expressed in the generic form
\be
\label{kramers}
r\simeq C\,e^{-\Delta V/T},
\ee
where $\Delta V$ is the energy barrier that has to be crossed, in line with the Arrhenius factor $e^{-\Delta V/T}$ first derived in the context of reaction rate theory \cite{Arrhenius:1889aa,Eyring:1935aa,Kramers:1940aa}. The prefactor $C$ depends on the dimensionality of the problem and is determined by the curvatures at the bottom and top of the potential wells \cite{Eyring:1935aa,Kramers:1940aa,Hanggi:1990aa}. There is a remarkable variety of activated processes in equilibrium that have been shown to follow the Kramers result, with only the detailed form of $C$ being model dependent \cite{Bovier:2015aa}. However, many real world systems, in particular biological ones, are intrinsically out-of-equilibrium due to the energy-dissipating active processes underlying their mechanical and dynamical properties \cite{MacKintosh:2010aa}. As a result, the effective fluctuations can be non-Gaussian, such that escape events are not governed by Kramers' result.

In this work, we show that Eq.~\eqref{kramers} is in fact a special case of a much more general expression that governs the escape behaviour in generic out-of-equilibrium systems that are driven by memoryless non-Gaussian fluctuations. Such fluctuations are ubiquitous in nature and have been shown to arise, e.g., in the dynamics of the cytoskeleton~\cite{ToyHeaSchMiz11}, intracellular transport \cite{Wang:2009aa,Chen:2015aa,AriTatTomMiz21}, and small tracer particles interacting with swimming microorganisms \cite{Leptos:2009aa,Krishnamurthy:2016aa,Kurihara:2017aa,Kanazawa:2020aa}. They also occur in technologically relevant nano-scale systems such as strongly coupled qubits \cite{Sung:2019aa} and Josephson junctions \cite{Ankerhold:2007aa,Huard:2007aa,Sukhorukov:2007aa,Grabert:2008aa}, and are often used in phenomenological descriptions of macroscopic dynamics, e.g., for animal foraging \cite{Sims:2008aa,Humphries:2010aa}, earthquake tremors \cite{Ditlevsen:1999ab}, and financial markets \cite{Cont:2003aa}. Memoryless non-Gaussian fluctuations are also implicit in many models of active matter, such as the widely studied run-and-tumble particles and other models, that exhibit, e.g., motility induced phase transitions \cite{TaiCat08,Fodor:2018ab}. As we discuss below, these systems can also be treated within our approach.

We present a unified framework for such noise processes based on path-integrals where exact results for both the escape rate and the optimal escape path are obtained. In this approach the general form of the Kramers rate is recovered but with $\Delta V$ replaced by an effective action that depends on both the detailed functional form of the potential and the noise parameters. We show that the effective action is, in fact, always lower than $\Delta V$ for symmetric noise, highlighting that non-Gaussian noise generically leads to exponential speed-ups of transition rates. This speed-up can be dramatic, as we show for a realistic swimmer model, where transition rates are increased by 25 orders of magnitude compared with the Gaussian case. We also discover that escape processes driven by non-Gaussian noise can exhibit large jumps in the most likely transition path, forming a separate universality class among such processes that is distinguished further by a non-Kramers form of the transition rate prefactor $C$, which we calculate explicitly. All our results are confirmed by numerical simulations.

\section{Model}

\subsection{Langevin dynamics driven by non-Gaussian noise}

We consider the time evolution of a single degree of freedom $q$, e.g.\ the position of a particle in one dimension, under the influence of a conservative force with potential $V$ as well as noise $\xi$, 
\be
\label{model}
\dot{q}(t)=-V'(q)+\xi(t)
\ee
In Eq.~\eqref{model}, we assume that all quantities are dimensionless, see \SI~Sec.~\ref{SI:dimless}.
Metastability occurs when $V(q)$ exhibits two or more sufficiently deep potential wells such that the particle is mostly confined to the bottom of one of the wells, with rare escape events to neighbouring wells induced by the noise \cite{Hanggi:1990aa}. We study the {\em rates} for such escape events, in a framework that can be extended to systems with many degrees of freedom and non-conservative forces (which, for the Gaussian case, have been studied in \cite{Bouchet:2016aa}). Key to our setup is that $\xi(t)$ contains not only the conventional (Langevin) Gaussian white noise, but an additional non-Gaussian contribution that breaks detailed balance:
\be
\xi(t)=\xi_{\rm G}(t)+\xi_{\rm NG}(t).
\ee
We take the latter as essentially the most general memoryless form of noise. This is Poissonian shot noise, which consists of a series of discrete `kicks' arriving at rate $\lambda_0$:
\be
\label{psn}
\xi_{\rm NG}(t)=\sum_{j=1}^{N_t}A_j\delta(t-t_j).
\ee
Here the times $t_j$ come from a Poisson process with rate $\lambda_0$, so that the total number $N_t$ of kicks within a time interval $[0,t]$ follows a Poisson distribution with mean $\lambda_0 t$. Each kick size (amplitude) $A_j$ is drawn identically and  independently from some distribution.

While models of the form Eqs.~\eqtoref{model}{psn} have been used on phenomenological grounds to model a large variety of processes in the sciences, recent work has also shown that the memoryless (or white) non-Gaussian noise of Eq.~\eqref{psn} arises as the result of systematic coarse-graining procedures in interacting particle systems. For example, in athermal granular systems coupled with a thermal reservoir, a system-size expansion shows that to leading order correlations with the environment can be neglected and white non-Gaussian fluctuations persist in addition to thermal Gaussian white noise \cite{Kanazawa:2015aa,Kanazawa:2015ab}. Moreover, the dynamics of a passive tracer interacting with active particles in suspension can be shown to universally reduce to a process with Poisson statistics at low densities \cite{Baule:2022aa}. Memoryless non-Gaussian fluctuations then arise in the long-time regime and are manifest, e.g., in the non-Gaussian features of the tracer's displacement distribution \cite{Kanazawa:2020aa}, see also Sec.~\ref{sec:numerics} below.

\begin{figure}
\centering
\includegraphics[height=2.75cm]{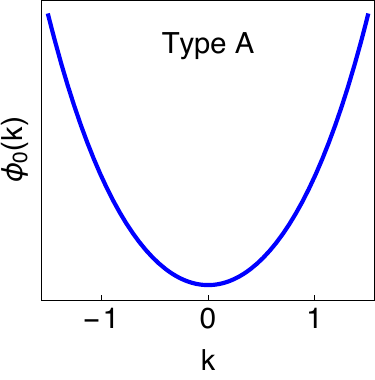}\;\;\;
\includegraphics[height=2.75cm]{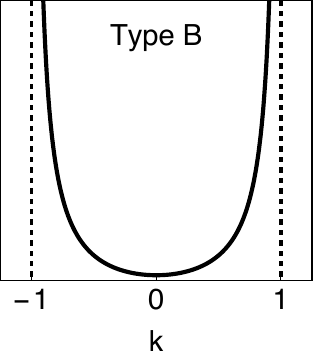}\;\;\;
\includegraphics[height=2.75cm]{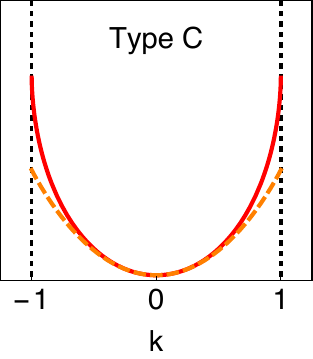}
\begin{tabular}{l|c|l|l}
Distribution  & Type & $\rho(x)$ & $\pphi(u)=\left< e^{u x}-u x-1\right>$  \\
\hline\hline
Constant & A & $\frac{\delta(x-1)}{2}+\frac{\delta(x+1)}{2}$ & $\cosh(u )-1$  \\
Exponential & B & $e^{-|x|}/4$ & $ u^2/[2(1-u^2)]$  \\
Gamma $\alpha$& C & $\frac{|x|^{-\alpha-1}e^{-|x|}}{2\Gamma(2-\alpha)}$ & $\frac{(1+ u)^\alpha+(1-u)^\alpha-2}{2\alpha(\alpha-1)}$ \\
\hline
Truncated $\phi$ & A &  & $\frac{1}{2} u^2 + b u^4$ 
\end{tabular}
\caption{\label{Fig_phis} We classify amplitude distributions $\rho$ into three types according to their moment generator $\pphi$. Type A: $\pphi$ is unbounded without singularities. Type B: $\pphi$ diverges upon approaching two singularities, taken as lying at $\pm1$. Type C: $\pphi$ is bounded with singularities at $\pm 1$ in higher derivatives. Examples for each type are given, normalized as $\int \D x\,x^2\rho(x)=1$. For the Gamma distribution we assume $0<\alpha<2$ and show the cases $\alpha=0.6$ (solid line) and $\alpha=1.6$ (dashed).}
\end{figure}

In order to investigate the dynamics of Eq.~\eqref{model}, we exploit the fact that the noise properties are captured by the cumulant generator (see \SI~Sec.~\ref{SI:pi})
\be
\label{cflevy}
\ln \big\langle e^{i\int_0^t \D s\,\xi(s)\uu(s)}\big\rangle =\int_0^t \D s\left[\frac{D_0}{2}(i\uu)^2+\lambda_0\phi(ig a_0)\right],
\ee
where $\phi$ is a moment generator defined as
\be
\label{phi}
\phi(u)= \int\D x\,\rho(x)\left( e^{u x}-u x-1\right)
\ee
The term $\frac{D_0}{2}(i\uu)^2$ in Eq.~\eqref{cflevy} represents the Gaussian white noise contribution, of variance $D_0$, while the second term $\lambdaorig\phi(iga_0)$ comes from the non-Gaussian kicks. Their amplitudes $A$ have the distribution $\rho(A/a_0)/a_0$ where the parameter $a_0$ sets the characteristic amplitude scale and $\rho(x)$ is a baseline distribution. The amplitude scale of this distribution $\rho(x)$ can then be fixed, which we do by imposing $\int \D x\,x^2\rho(x)=1$. All noise statistics can be obtained from Eq.~\eqref{cflevy}, e.g.\ $\langle \xi(t)\xi(t')\rangle=(D_0+\lambdaorig \aorig^2)\delta(t-t')$. Eq.~\eqref{cflevy} is in fact the most general form of the cumulant generator for a (zero mean) noise process $\xi$ that is stationary and uncorrelated in time\footnote{This is also known as L\'evy noise, and defined technically as the derivative of a process with independent stationary increments~\cite{Cont:2003aa}. Our only restriction on this is the finiteness of $\int\D x\, x^2\rho(x)$, to allow us to assign a scale to the noise variance.}. In this form the setting also covers cases where $\rho$ is not normalizable, e.g.\ when it has a power law divergence $\rho(x)\propto |x|^{-\alpha-1}$ for small $x$ \cite{Koponen:1995aa} with $0<\alpha<2$. We focus in the following on symmetric noise with $\rho(x)=\rho(-x)$. Our analysis will show that escape properties depend crucially on the form of $\rho$; in fact we will be able to classify amplitude distributions $\rho$ into three different types A,B,C as illustrated in Fig.~\ref{Fig_phis}.

Eqs.~\eqtoref{model}{phi} unify the description of non-Gaussian noise-induced activation studied previously, both analytically and numerically, for a range of special cases such as kicks with exponentially distributed \cite{Van-den-Broeck:1984aa,Sancho:1985aa,Masoliver:1987aa,Hernandez-Garcia:1987aa,Porra:1993aa,Laio:2001aa,Gera:2021ve} or constant amplitudes \cite{Billings:2008aa,Dykman:2010aa,Khovanov:2014aa}, and L\'evy flights \cite{Ditlevsen:1999aa,Bao:2005aa,Chechkin:2005aa,Dybiec:2006aa,Dybiec:2007aa,Chechkin:2007aa,Chen:2011aa,Gao:2014aa}. We also include in our considerations the form of $\phi$ obtained by expanding to the first non-Gaussian order (cubic in general, quartic in our symmetric case). This widely used approximation scheme corresponds to artificially setting to zero all higher cumulants of the noise amplitude distribution 
\cite{Ankerhold:2007aa,Huard:2007aa,Sukhorukov:2007aa,Grabert:2008aa,Khovanov:2014aa,Li:2020tb} and we will see that it can lead to qualitatively incorrect predictions. Our framework will also allow us to recover rigorous mathematical results on the dominant scaling of the escape rate for non-Gaussian noise for a specific weak-noise regime \cite{Imkeller:2009aa,Imkeller:2010aa}.

\subsection{Path-integral framework}

Our analysis of non-Gaussian escape rates is based on a path integral framework. As in the seminal Kramers escape rate calculation for Gaussian noise, we will consider a weak-noise regime. Fluctuations around the most likely escape path from one metastable state to another are then small and the typical path can be obtained by minimizing a stochastic action $S[q]$ w.r.t.\ to paths $q(s)$. The key technical steps in extending this approach to the non-Gaussian case with cumulant generator given by Eq.~\eqref{cflevy} are (see \SI~Sec.~\ref{SI:pi}): (1.) Following the Martin-Siggia-Rose formalism, the transition probability of the escape process is expressed as an integral over paths $q(s),g(s)$, where $g$ is an auxiliary field conjugate to the noise. (2.) We rescale the noise parameters by a dimensionless scaling parameter $\epsilon$ as 
\be
\label{rescaling}
D_0=D\,\epsilon,\qquad \lambdaorig=\lambdatilde/\epsilon,\qquad \aorig= \atilde\,\epsilon
\label{epsilon_scaling}
\ee
The variance of the noise $\xi$ is then $D_0+\lambdaorig \aorig^2=(D+\lambdatilde\atilde^2)\epsilon\propto \epsilon$ so that the weak noise limit is $\epsilon\to0$. While Eq.~\eqref{rescaling} may appear somewhat specific, it represents in fact a generic weak-noise regime that preserves all details of the non-Gaussian noise distribution $\rho$ for small $\epsilon$, not just the leading non-Gaussian cumulants as considered in \cite{Ankerhold:2007aa,Huard:2007aa,Sukhorukov:2007aa,Grabert:2008aa,Khovanov:2014aa} (see the discussion in \SI~Sec.~\ref{App_Sec_rescaling}). We also emphasize that our final results can be converted again into the original noise parameters $D_0,\lambdaorig,\aorig$ or into another weak-noise regime, see Sec.~\ref{Sec_special}, highlighting the generality of our approach.

(3.) The auxiliary field $g$ can be integrated out by a saddle-point method for $\epsilon\to 0$. The net result for the path probability takes the large-deviation form
\be
\label{pathprob3}
P[q]\propto e^{-\int_0^t\D s\,\lagr(\dot q+V'(q))/\epsilon},
\ee
with the Onsager-Machlup-like Lagrangian $\lagr$ written only in terms of the physical paths $q(s)$. We find that $\lagr(\cdot)$ is given by the Legendre transform $\lagr(\f)=\max_\mom [\mom\f - \psi(\mom)]$ of
\be
\label{psi}
\psi(\mom)=D\mom^2/2+\lambdatilde\phi(\atilde\mom).
\ee
One can check that this result remains valid even when $\phi$ has singularities on the real axis; the maximum then has to be taken over the non-singular range. In the example cases shown in Fig.~\ref{Fig_phis} such singularities occur for the Gamma and exponential noise amplitude distributions. In contrast, distributions with tails decaying faster than exponentially do not produce singularities in $\phi$; see the constant modulus example in Fig.~\ref{Fig_phis}. We note for later that $\phi$ is convex and therefore so are $\psi$ and the Lagrangian $\lagr$ as its Legendre transform. For our symmetric noise distributions, all three functions are also symmetric and thus have their global minimum at vanishing argument. The symmetry further ensures that all odd moments of $x$ vanish while the even ones are positive, which from Eq.~\eqref{phi} implies the lower bound $\phi(u) \geq u^2/2$ and hence a similar bound $\psi(\mom)\geq (D+\lambdatilde \atilde^2)\mom^2/2$.

\section{Escape from a metastable state}

\subsection{Effective action}

Let us now consider escape from a metastable state $q_a$, located at the minimum of the metastable basin of $V$, across the top of the nearest potential barrier at $q_b>q_a$. For Gaussian noise, the path integral solution of this problem \cite{Caroli:1981aa,Weiss:1982aa,Bray:1989aa} is analogous to the quantum mechanical tunneling problem treated in a semiclassical approximation \cite{Kleinert:2009aa} and gives the dominant scaling of the escape rate $r$ for small $D$ as $r\cong C\,e^{-\Smin/D}$. In our general non-Gaussian case the equivalent form can be deduced from the theory of large deviations \cite{Freidlin:1998aa} for $\epsilon\to 0$ with the effective energy barrier given by the minimum action
\be
\label{action}
\Smin=\lim_{t\to\infty} \min_{[q]}\int_0^t\D s\,\lagr(\dot{q}+V'(q)).
\ee
The minimum is over all paths with $q(0)=q_a$, $q(t)=q_b$, and the resulting optimal path (also called `instanton' or `excitation path') gives the typical escape trajectory for small $\epsilon$. To make progress in determining $\Smin$, one can think of any $q(s)$ as a path in the $(q,v)$-plane, with $v=\dot q$. Then the action reads $\int \D q\, \lagr(v+ V'(q))/|v|$ and for each $q$ we can find $v=\dot q$ simply as the minimum\footnote{We do not need to enforce the total time constraint $t=\int \D q/|v|$ as the minimal action path is obtained for $t\to\infty$, which is automatically fulfilled since the integral for $t$ diverges at both ends for paths between stationary points of $V$.} of $\lagr(v+ V'(q))/|v|$. The trivial global minimum is $v=-V'(q)$, which describes deterministic relaxation. For an excitation from $q_a$ to $q_b>q_a$, on the other hand, we have $V'>0$ and need $v>0$. If -- and this is an important restriction as we will show -- the minimum of $\lagr(v+ V'(q))/v$ occurs at finite $v$, it obeys $\lagr(v+ V'(q))=v\lagr'(v+ V'(q))$. This condition, together with the fact that $\lagr$ is the Legendre transform of $\psi$, i.e., $\lagr'(f)=\mom^*$ with $\mom^*={\rm argmax}_\mom [\mom\f - \psi(\mom)]$
yields for the minimum action the simple result
\be
\label{Sgen}
\Smin=\int_{q_a}^{q_b} \D q\,\mom^*(V'(q)),
\ee
where $\mom^*(V')$ is determined from
\be
 V'(q) &=& \psi(\mom^*)/\mom^*.
\label{k_def}
\ee
This expression is just our minimum condition $\lagr(f)=v\mom^*$ rewritten using $\lagr(f)=\mom^* f-\psi(\mom^*)$ and $f=v+V'$.
The inverse Legendre transform relation $\psi'(\mom^*)=f$ yields further
\be
\label{instanton}
v= \dot{q}=\psi'(\mom^*)-V'(q).
\ee
Together with Eq.~\eqref{k_def} this defines a velocity function $\dot q=\Xi(V'(q))$ that characterizes the shape of the instanton.

By comparing Eq.~\eqref{Sgen} with the classical mechanics result $\partial S/\partial q=p$ one sees that our $\mom^*$ plays exactly the role of momentum, while the minimization condition $\mom^*V'(q)=\psi(\mom^*)$ corresponds to the well-known condition that the Hamiltonian $\ham=\mom^* \dot q-\lagr = -\mom^* V'(q)+\psi(\mom^*)$ must vanish on minimum action paths of duration $t\to\infty$ \cite{Billings:2008aa,Gera:2021ve}. However, we will discover below that minimal action paths can in certain cases contain jumps, in which case the criterion $\ham=0$ ceases to be applicable because $\dot q$ becomes undefined. Our approach of minimizing $\lagr(v+V'(q))/|v|$ will continue to be valid, on the other hand.

\subsection{Gaussian vs non-Gaussian escape}

\begin{figure*}
\centering
\includegraphics[height=4cm]{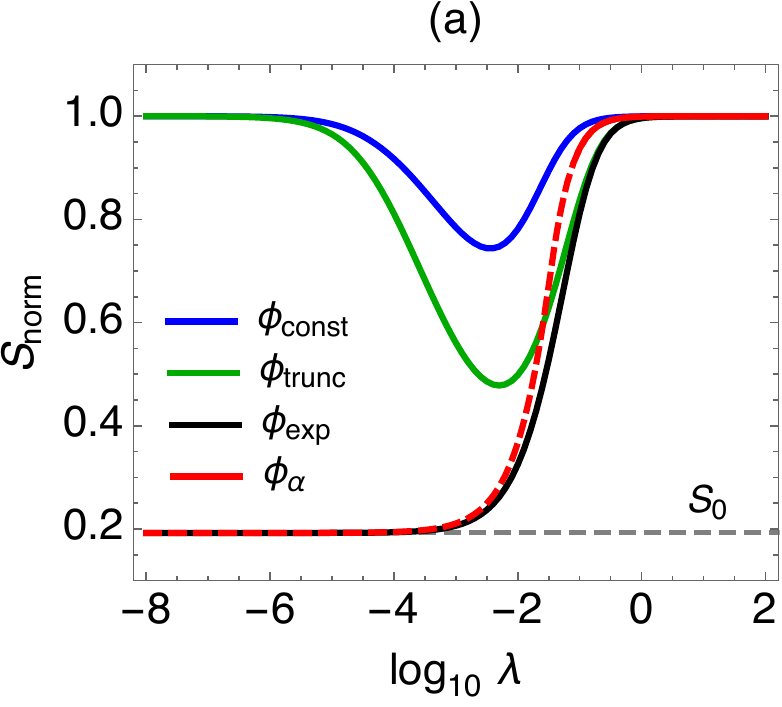}\hspace{0cm}
\includegraphics[height=4cm]{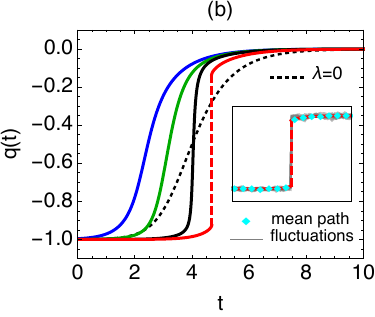}\hspace{0cm}
\includegraphics[height=4cm]{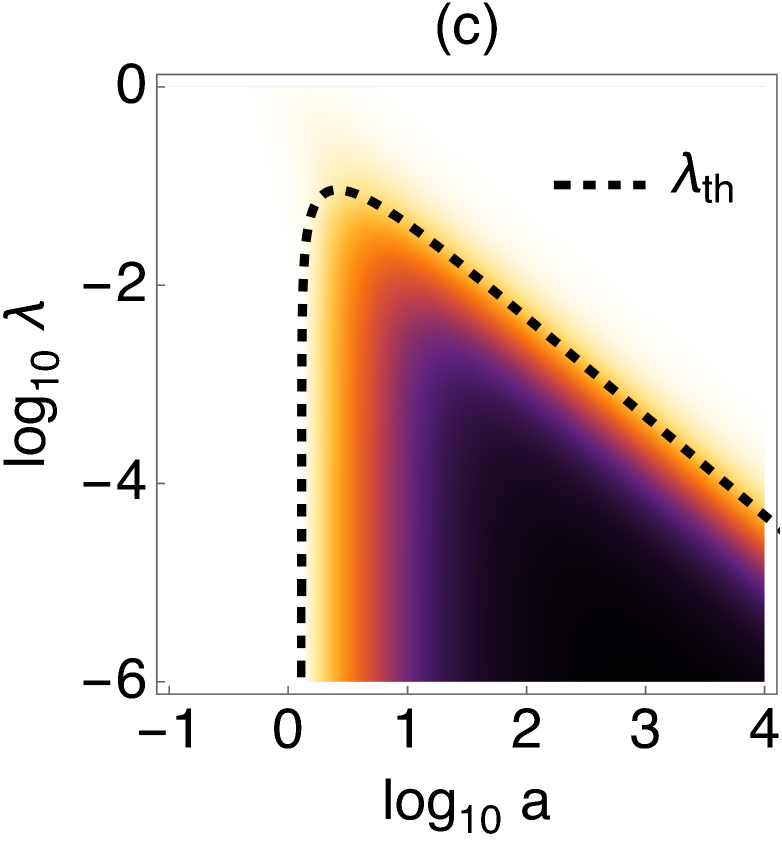}\hspace{0cm}
\includegraphics[height=4cm]{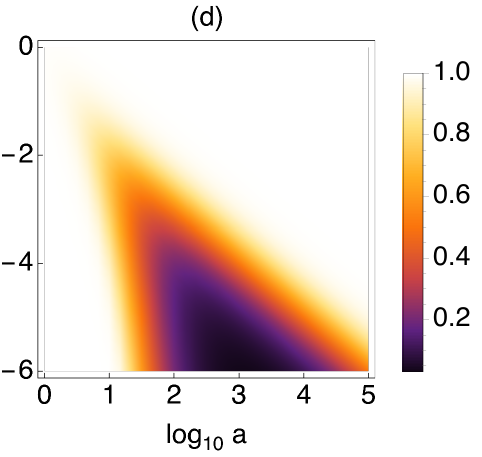}
\caption{\label{Fig_action} (a) The normalized action $S_{\rm norm}=\Smin/\Sg$ for the different $\pphi$ of Fig.~\ref{Fig_phis} ($\atilde=10$, $\alpha=0.8$, $b=1/2$, $D=1$) and the potential $V(q)=q^4/4-q^2/2$. Noise amplitude distributions of type A ($\pphi_{\rm{const}},\phi_{\rm{trunc}}$) recover the Gaussian value $S_{\rm norm}=1$ as $\lambdatilde\to 0$. For type B and C amplitudes ($\pphi_{\rm{exp}},\pphi_{\alpha}$), $S_{\rm norm}$ decreases monotonically as $\lambdatilde\to 0$ and converges to a nontrivial limit $S_0$, Eq.~\eqref{S0}. The action for $\pphi_{\alpha}$ corresponds to an escape path with a discontinuous jump when $\lambdatilde<\lambdatildec$ (red dashed line).  (b) Instanton escape paths for the different $\pphi$ showing a rapid motion from the initial minimum to the barrier; for $\phi_{\alpha}$ the instanton has a jump section. Colors, potential, and parameters as in (a) (apart from $\alpha=1.2$ and $\lambdatilde=0.01$). For non-Gaussian noise the time-reversal symmetry between excitation and relaxation paths is broken, seen here by the difference with the slower $\lambdatilde=0$ instanton of the Gaussian dynamics (dotted line). Inset: Mean path sampled numerically from the path weight for $\epsilon=0.01$ compared with theory for $\pphi_{\alpha}$, confirming the jump. (c,d) Colour maps of $S_{\rm norm}$ for (c) $\pphi_{\alpha}$ with $\alpha=0.8$ and (d) $\pphi_{\rm{const}}$. The dashed line is the phase boundary $\lambdatilde_{\rm th}(\atilde)$ separating regions with smooth ($\lambdatilde\ge\lambdatilde_{\rm th}$) and jump ($\lambdatilde<\lambdatilde_{\rm th}$) instantons.}
\end{figure*}

Analysing the effective energy barrier $\Smin$ for arbitrary non-Gaussian noise types yields striking differences with the Gaussian case summarized as follows:
(i) Non-Gaussian noise always generates larger escape rates, i.e.\ it is at least as efficient as Gaussian noise: $\Smin< 2\,\Delta V/(D+\lambdatilde \atilde^2)\equiv \Sg$ for any distribution of type A, B or C, see Figs.~\ref{Fig_action}a,b. The reference value $\Sg$ here is the activation barrier that results when the non-Gaussian noise is replaced by Gaussian noise of the same variance, corresponding to the truncation of the Taylor expansion of $\pphi(k)$ after the quadratic term. Because $\Smin$ enters the escape rate as $\exp(-\Smin/\epsilon)$, non-Gaussian noise thus offers {\em exponential} speed-ups. (ii) Remarkably, for amplitude distributions of types B and C even noise of infinitesimal intensity $\lambdatilde\to 0$ yields a value of $\Smin$ considerably smaller than $\Sg$, indicating a singular limit. (iii) Optimal escape paths have the characteristic instanton shape, with the particle moving rapidly from the initial minimum to the transition state at the top of the barrier, but the shape varies with $\pphi$. This contrasts with the Gaussian noise case, where excitation paths are simply the time-reverse of deterministic relaxation paths (Fig.~\ref{Fig_action}). (iv) For type C amplitude distributions we identify an entire region in the $(\atilde,\lambdatilde)$ parameter plane where the escape paths contain a {\em discontinuous jump} (Fig.~\ref{Fig_action}). Note that the behaviours (ii) and (iv) cannot be reproduced with any cumulant truncation, as this effectively produces a type A form of $\pphi(k)$.

We proceed to explain all of these observations on the basis of the properties of the noise amplitude moment generator $\pphi$. Firstly we saw above that $\psi(\mom)\geq (D+\lambdatilde \atilde^2)\mom^2/2$, which implies from Eq.~\eqref{k_def} that $\mom^*\leq 2V'/(D+\lambdatilde \atilde^2)$. With Eq.~\eqref{Sgen} the reduction (i) of the effective barrier, $\Smin < \Sg$, follows directly. 

To analyse the limit of small $\lambdatilde$ we consider the solutions of Eq.~\eqref{k_def}, which using Eq.~\eqref{psi} can be cast in the form
\be
\label{lambdaeq}
V'(q)=\frac{D}{2}\,\mom^*+\lambdatilde \frac{\pphi(\atilde\mom^*)}{\mom^*}.
\ee
Rewriting further one can show that in order to see strongly non-Gaussian behaviour the noise amplitude has to lie in the range $1\ll \atilde \ll 1/\lambdatilde$ (see \SI~Sec.~\ref{SI:escape}), which in turn requires $\lambdatilde \ll 1$. Considering accordingly $\lambdatilde \to 0$ for fixed $\atilde$, the last term in Eq.~\eqref{lambdaeq} disappears, suggesting that $\mom^*=2V'/D$, which yields Gaussian behaviour. This argument always works for amplitudes of type A, while for type B it only holds if $\mom^*=2V'/D$ remains smaller than the singularity in $\pphi(\atilde\mom)$ at $1/\atilde$, i.e., when $2V'/D<1/\atilde$. If instead $2V'/D>1/\atilde$, the solution of Eq.~\eqref{lambdaeq} approaches $\mom^*=1/\atilde$ for $\lambdatilde\to 0$, since $\pphi(\atilde\mom)$ diverges for $\mom\to 1/a$ and the last term in Eq.~\eqref{lambdaeq} eventually becomes dominant. Overall, one therefore obtains $\mom^*(V')\to\min(2V'/D,1/\atilde)$ and the effective energy barrier $\Smin\to S_0$ where from Eq.~\eqref{Sgen}
\be
\label{S0}
S_0 = \int_{q_a}^{q_b} \D q \min(2V'(q)/D,1/\atilde)
\ee
The value $S_0$ that is approached as $\lambdatilde\to 0$ lies {\em below} $\Sg$ for $1/\atilde<2\max_q V'(q)/D$, making the limit discontinuous\footnote{The discontinuity is possible as we have implicitly taken the limit $\epsilon\to 0$, where the unscaled rate $\lambda_0$ is large for any $\lambdatilde>0$ from Eq.~\eqref{epsilon_scaling}.} (see Fig.~\ref{Fig_action}. Since for type B noise $\pphi'(\atilde\mom)$ and thus $\psi'(\mom)$ diverge for $\mom\to 1/\atilde$, Eq.~\eqref{instanton} tells us that the velocity $v$ becomes very large on the sections of the instanton with $1/\atilde<2V'/D$, so that the path in that region becomes closer and closer to a discontinuous jump as $\lambdatilde$ decreases. On the other hand, on the sections with $1/\atilde>2V'/D$, Eq.~\eqref{instanton} yields the Gaussian shape.

Remarkably, for type C, the boundedness of $\pphi$ implies that Eq.~\eqref{lambdaeq} will not have a solution when $V'$ lies above a threshold $\Vc=\max_\mom \psi(\mom)/\mom=D/(2\atilde)+\lambdatilde \atilde\pphi(1)$. This condition is met on at least part of the instanton when $\lambdatilde<\lambdatildec=[\max_q V'(q)-D/(2\atilde)]/[\atilde\pphi(1)]$. In the range of $q$ where $V'(q)>\Vc$ our approach shows its key benefit over the standard Euler-Lagrange equations or the criterion $\ham=0$, neither of which have solutions in this regime because $\dot q$ becomes undefined: one can check here that $\lagr(v+V'(q))/v$ is monotonically decreasing for $v>0$, reaching the limit $1/\atilde$ for $v\to\infty$: the optimal velocity is infinite, $\Xi(V'(q))=\infty$. This implies that there must be a jump in the optimal path whenever $\lambdatilde<\lambdatildec$. To the action this jump contributes $\int \D q/a=\Delta q/\atilde$ where the integral covers the relevant $q$-range and gives the length $\Delta q$ of the jump. The contribution of the rest of the path has to be found by solving Eqs.~\eqqref{k_def}{instanton} as before, which produces the Gaussian shape for $\lambdatilde \ll 1$. The condition $\lambdatilde<\lambdatildec$ maps out a dynamical phase diagram in the $(\atilde,\lambdatilde)$ plane separating jump and no-jump escape behaviours (see Fig.~\ref{Fig_action}c).

Since the threshold $\Vc\to D/(2\atilde)$ for $\lambdatilde \ll 1$, the escape behaviour for noise amplitude distributions of type B,C becomes identical in this regime: the instanton consist of initial and final segments of time-reversed relaxations, connected by a jump, and the resulting action is $S_0$, Eq.~\eqref{S0}. We remark that the class of amplitude distributions with this property can be characterized generally as distributions with exponentially decaying tails, i.e.\ of the form $\rho(x)= c(x)e^{-|x|}$, with $\lim_{x\to\pm\infty}\ln(c(x))/x=0$. These two conditions are sufficient for the existence of a singularity at $\pphi(1)$, see Eq.~\eqref{phi}. Jump instantons at finite $\lambdatilde$ as in type C appear when, in addition, the condition $\int_1^\infty \D x\,c(x)<\infty$ is satisfied, since then $\pphi(1)$ is finite.

\subsection{Special cases}
\label{Sec_special}

Our general solution in Eqs.~\eqtoref{Sgen}{instanton} reproduces existing results in the literature for specific amplitude distributions. As a sanity check, we find in the Gaussian case ($\lambdatilde=0$) $\psi(\mom)=D\mom^2/2$; thus $\mom^*=2V'/D$, which with Eq.~\eqref{Sgen} and the Einstein relation $D\epsilon=D_0=2T$ recovers the van't Hoff--Arrhenius scaling $\sim e^{-\Delta V/T}$ of the escape rate. The instanton obeys $\dot{q}=V'(q)$ from Eq.~\eqref{instanton}, which as expected for Gaussian noise is the time reverse of a noise-free deterministic relaxation path \cite{Caroli:1981aa,Weiss:1982aa,Bray:1989aa}. For escape driven by one-sided exponentially distributed amplitudes without a Gaussian component, we have $\phi(u)=u^2/(2(1-u))$ and solving Eq.~\eqref{k_def} for $k^*$ yields $k^*=2V'/(\lambda a^2+2aV')$ as obtained in \cite{Van-den-Broeck:1984aa,Sancho:1985aa,Porra:1993aa}. We likewise recover analytical results for the effective action derived for one-sided constant and two-sided exponentially distributed amplitudes \cite{Billings:2008aa,Gera:2021ve}, see \SI~Sec.~\ref{SI:escape}.

Rigorous mathematical results for the escape rates of Eqs.~\eqtoref{model}{phi} have been obtained in \cite{Imkeller:2009aa,Imkeller:2010aa} for a different scaling regime of the noise parameters. Remarkably, our large deviation approach is able to recover these results for those amplitude distributions for which $\phi$ from Eq.~\eqref{phi} is well-defined. Instead of Eq.~\eqref{rescaling}, the parameter scaling adopted in \cite{Imkeller:2009aa,Imkeller:2010aa} is given by
\be
\label{rescaling2}
D_0=\epsilon^2,\qquad \lambdaorig=1,\qquad \aorig= \epsilon,
\ee
which leads to a weak-noise regime with a constant rate of non-Gaussian noise kicks and intensity $D_0+\lambdaorig \aorig^2= \epsilon^2$. We can retrieve this scaling by setting $D=\lambda=\epsilon'$ and $a=1$ {\it after} the rescaling in Eq.~\eqref{rescaling} that leads to the large deviation form of the action. We then take $\epsilon'$ as small and identify $\epsilon'=\epsilon$ at the end. Now for $\epsilon'\ll 1$ the solutions of Eq.~\eqref{lambdaeq} satisfy
\be
\label{lambdaeq2}
V'(q)=\frac{\epsilon'}{2}k^* +\epsilon' \frac{\phi(k^*)}{k^*}\approx \epsilon' \frac{\phi(k^*)}{k^*},
\ee
since $k^*$ will become large for small $\epsilon'$ and $\phi(u)$ increases at least exponentially for large $\mom^*$. Two classes of amplitude distributions discussed in \cite{Imkeller:2009aa,Imkeller:2010aa} are bounded amplitudes such as the constant amplitudes of Fig.~\ref{Fig_phis}, and amplitude distributions with super-exponentially decaying tails, $\rho(x)\propto \exp(-x^\gamma)$ with $\gamma>1$. For the former we have $\phi(u)\sim e^{bu}$ for $u\gg 1$, where $b$ is the upper bound, and for the latter $\phi(u)\sim \exp\left[(\gamma-1)\left({u}/{\gamma}\right)^{\gamma/(\gamma-1)}\right]$. Determining then the asymptotic solutions of Eq.~\eqref{lambdaeq2} for $\epsilon'\ll 1$ and substituting into Eq.~\eqref{Sgen} with $\epsilon'=\epsilon$ yields the dominant terms in the effective action for $\epsilon\to 0$ as
\be
\label{im1}
S_{\rm min}\approx (q_b-q_a)|\ln\,\epsilon |
\ee
for bounded amplitudes and
\be
\label{im2}
S_{\rm min}\approx (q_b-q_a)\gamma(\gamma-1)^{(1-\gamma)/\gamma}|\ln\,\epsilon |^{(\gamma-1)/\gamma}
\ee
for amplitude distributions with super-exponentially decaying tails. Eqs.~\eqqref{im1}{im2} are precisely the results obtained\footnote{We note that our approach is not able to reproduce the corresponding expressions for amplitude distributions that decay with power-law or sub-exponential tails calculated in \cite{Imkeller:2009aa,Imkeller:2010aa} since in these cases $\phi(u)$ is undefined for any nonzero real $u$.} in \cite{Imkeller:2009aa,Imkeller:2010aa} for $r\propto e^{-\Smin/\epsilon}$ and $\epsilon\ll 1$.

\subsection{Prefactor}

The effects discussed above relate to the exponential term in the rate of escape processes $r\simeq C\exp(-\Smin/\epsilon)$, with non-Gaussian noise producing exponential speed-ups by reducing $\Smin$. We have also studied the prefactor $C$, to see whether this modifies the results. Recent work has shown that $C$ can be determined by solving matrix Riccati equations, which is particularly suitable for numerical evaluations \cite{Grafke:2021aa,Bouchet:2021aa}. Analytical expressions for $C$ have previously been obtained e.g.\ by calculating the fluctuation determinant in the path integral approach \cite{Caroli:1981aa,Weiss:1982aa,Luckock:1990aa} or by determining steady state solutions \cite{Kramers:1940aa,Hanggi:1990aa} of the Fokker-Planck equation associated with Eq.~\eqref{model}, augmented by an injection term near $q_a$. We have used both these methods to confirm that in the regime where the excitation path is smooth, the prefactor is exactly the {\em same} as in the Gaussian case, i.e.\ given by the Eyring-Kramers expression $C=\sqrt{V''(q_a)|V''(q_b)|}/(2\pi)$ \cite{Sollich:2022aa}, as observed previously for special cases of our noise \cite{Van-den-Broeck:1984aa,Gera:2021ve}.

However, $C$ is modified when the excitation path has a jump section. The path integral method breaks down here because the eigenfunction expansion of the relevant fluctuation operator becomes ill-defined. However, determining the flux over the barrier in steady state remains feasible. We report the technically non-trivial calculation elsewhere \cite{Sollich:2022aa}. The result applies generally to noise distributions $\rho(x)=c(x)e^{-|x|}$ with exponential cutoff and power law tails, $c(x)\simeq c_\alpha x^{-\alpha-1}$ for $x\gg 1$. We find
\be
C= c_\alpha\epsilon^{\alpha} \lambdatilde [(q_+-q_-)/\atilde]^{-\alpha-1}\left(\frac{V''(q_a)|V''(q_b)|}{V''(q_-)|V''(q_+)|}\right)^{1/2}
\label{eq:C}
\ee
if the jump is from $q_-$ to $q_+$. The key observation here is that while the prefactor is no longer independent of $\epsilon$, its power law variation $\epsilon^\alpha$ is much weaker than the exponential $\exp(-\Smin/\epsilon)$. For small $\epsilon$ non-Gaussian noise therefore still generates vastly faster escapes from metastable states than Gaussian noise of the same variance. We also observe in Eq.~\eqref{eq:C} that the (scaled) rate $\lambdatilde$ of the non-Gaussian noise enters as a prefactor, demonstrating that the escape dynamics is largely controlled by non-Gaussian effects. These must then disappear for $\lambdatilde=0$ or more precisely, by comparing with the Kramers rate, when $\lambdatilde$ becomes of $\mathcal{O}(e^{-(\Sg-S_0)/\epsilon})$. The final factor in Eq.~\eqref{eq:C} contains the curvature information from the Kramers prefactor but effectively corrects this by the relevant curvatures at the beginning and end of the jump. Note that the remaining factors can be written as $\lambdatilde \epsilon^{-1} c((q_+-q_-)/(\epsilon\atilde))$ using the large $x$-behaviour of $c(x)$, and in that form should be generic for other, less than exponentially varying, forms of $c(x)$ that produce discontinuous excitation paths. This contribution to $C$ is essentially the probability of receiving a noise ``kick'' that will perform the required jump. The exponential  factor $e^{-|x|}=e^{-(q_+-q_-)/(\epsilon \atilde)}$ from $\rho(x)$ that should also appear here is accounted for in the action $\Smin$ and is exactly the jump contribution to $\Smin$ we identified earlier.

\begin{figure}
\centering
\includegraphics[height=3.75cm]{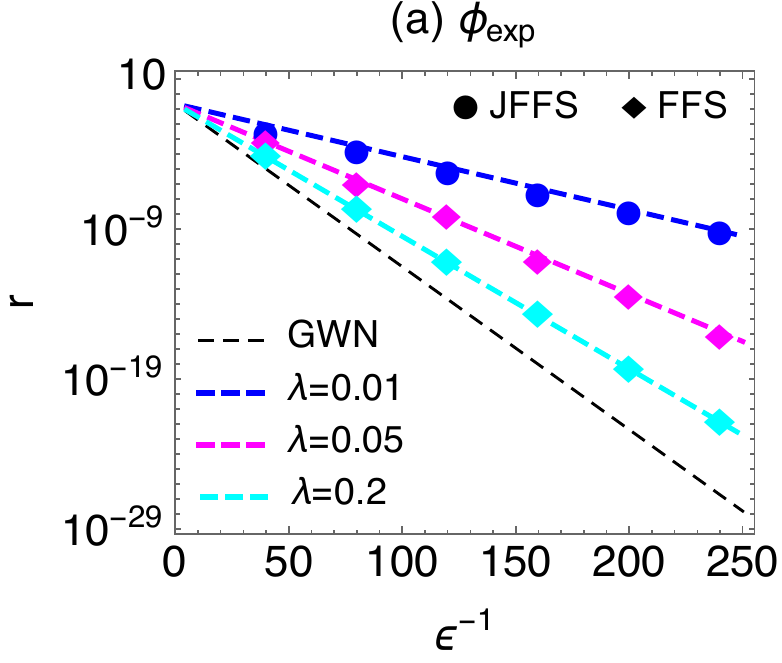}\hspace{-0.4cm}
\includegraphics[height=3.75cm]{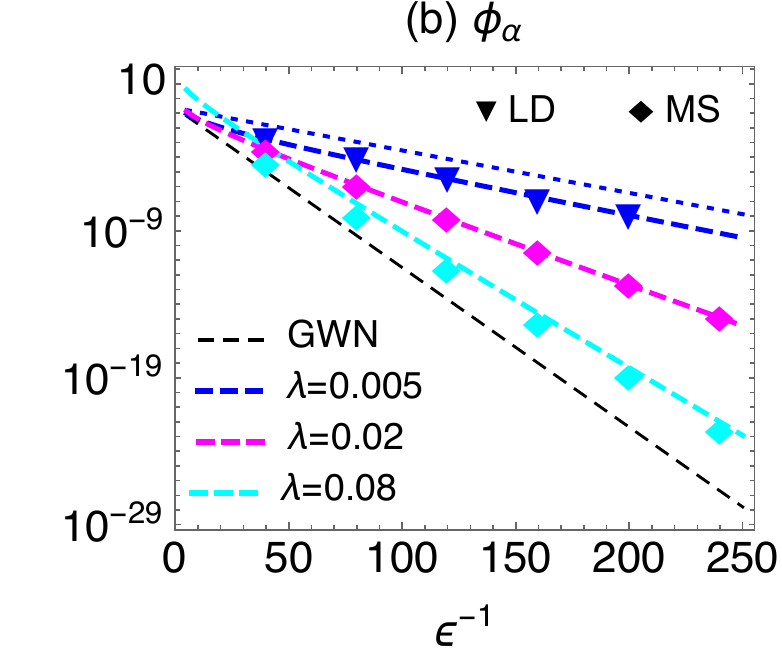}
\caption{\label{Fig_numerics} Comparison of the theoretical predictions (lines) for the escape rate $r$ and results from numerical simulations. We have employed forward-flux sampling (FFS), jumpy forward-flux sampling (JFFS), direct Langevin simulations (LD), and a numerical solution of the Master equation (MS). The potential is $V(q)=q^4/4-q^2/2$, $D=1$, and we have set the rescaled noise intensity $\lambda a^2$ of the non-Gaussian component also to unity, i.e., $a=1/\sqrt{\lambda}$, which leaves $\lambda$ as the only free parameter. Dashed black line: escape rates for purely Gaussian noise of the same noise intensity (GWN), highlighting exponential speed-ups due to non-Gaussian effects. (a) PSN with exponentially distributed amplitudes, leading to smooth instantons. (b) Gamma noise with $\alpha=0.8$, leading to jump instantons for $\lambda=0.005$ and $\lambda=0.02$; the predicted jump prefactor (\ref{eq:C}) clearly gives a better description of the data than the Eyring-Kramers prefactor (dotted line for $\lambda=0.005$). In both (a) and (b) the largest escape rates are achieved for $\lambda\to 0$, see also Fig.~\ref{Fig_action}(a).}
\end{figure}

\subsection{Comparison with simulations}
\label{sec:numerics}

To check our theoretical predictions, we implemented different simulation algorithms to determine the escape rates numerically. The Langevin dynamics can be simulated with standard methods based on an Euler discretization of Eq.~\eqref{model} \cite{Cont:2003aa}, but escape events become exceedingly rare as $\epsilon\to 0$ and measuring very small rates thus requires suitable rare-event sampling algorithms. In the parameter regime in which the instanton is smooth, we have used forward-flux sampling (FFS) \cite{Allen:2009aa} and jumpy FFS \cite{Haji-Akbari:2018aa} to confirm the theoretical predictions, see Fig.~\ref{Fig_numerics}a, reaching rates as small as $r\approx 10^{-22}$. These methods are not applicable when the instanton has jumps, since for escape events with jumps the partitioning of the coordinate space into neighbouring bins as used in FFS becomes meaningless. In the jump regime, we thus used direct Langevin simulations (DL) and, in order to reach smaller rates, a numerical solution of the Master equation associated with Eq.~\eqref{model} (MS), which confirm our theory and demonstrate in particular the validity of the prefactor Eq.~\eqref{eq:C}, see Fig.~\ref{Fig_numerics}b.

Fig.~\ref{Fig_numerics} highlights the exponential increase of escape rates due to non-Gaussian noise, which achieves speed-ups of up to 20 orders of magnitude for the same noise intensity. Conversely, this dramatic difference implies that assessing the effect of fluctuations on transition rates based on their variance alone is unreliable and can drastically underestimate the true transition rate. To elucidate this point further we investigated a realistic non-Gaussian noise-driven system to compare our predictions with the Kramers theory. We simulate non-interacting swimmers in a three dimensional volume interacting with a passive tracer particle via a truncated dipolar force that describes the hydrodynamic interaction in the far-flow field regime at low Reynolds numbers \cite{Kanazawa:2020aa}. As shown in \cite{Kanazawa:2020aa}, the stochastic motion of the tracer is effectively driven by non-Gaussian noise described by Eqs.~\eqqref{model}{phi}, provided the dynamics is observed on sufficiently long time scales. Trapping the tracer in the double well potential $V(q)=V_0 \left[(q/q_0)^4/4-(q/q_0)^2/2\right]$, we measure the escape rate as
\be
\label{rvalue}
r\approx 2\cdot 10^{-6}
\ee
for $V_0=5\cdot10^{-6}$ and $q_0=25$. These parameters have been adjusted such that escape times are short enough to be measurable, but also sufficiently long 
to achieve the Markovian regime of the tracer dynamics (all remaining parameters are set as in \cite{Kanazawa:2020aa}). Calculating the escape rate with Eqs.~\eqqqref{psi}{Sgen}{k_def} where $\phi$ is fitted from the empirical tracer displacement statistics yields a rate of $r\approx 6\cdot 10^{-4}$. While this differs by two orders of magnitude from the measured rate in Eq.~\eqref{rvalue},  Kramers rate theory based on the diffusion coefficient of the tracer would give $r\approx 5\cdot 10^{-32}$(!), again emphasizing that ignoring the non-Gaussian characteristics of the tracer can lead to dramatically inaccurate predictions.

\section{Discussion}

Our results demonstrate that non-Gaussian noise can induce qualitatively very different escape behaviours. The instantons with jump section, occurring within the jump phase shown in Fig.~\ref{Fig_action}c indicate an escape strategy that is fundamentally different from the one we find in thermal equilibrium systems: instead of completing the entire escape using a rare sequence of small fluctuations, the system prefers to wait for a single rare fluctuation that is large enough to carry it across the steepest section of the potential barrier. Remarkably, the prefactor $C$ highlights the existence of two universality classes associated with these two types of escape: the Kramers prefactor, which also applies to non-Gaussian noise in the parameter range where the escape path is smooth; and Eq.~\eqref{eq:C} that governs the jump escape. 

The theoretical analysis shows that the exponential speed-up of transition rates persists and becomes even more pronounced in the regime $\lambdatilde\to 0$, i.e.\ when the non-Gaussian contribution in Eq.~\eqref{phi} is vanishingly small, see Fig.~\ref{Fig_action}a. It might be possible to exploit this effect to optimize switching behaviour in artificial systems driven by non-Gaussian noise such as colloids interacting with an active microbial heat bath on which thermodynamic cycles can be imposed \cite{Krishnamurthy:2016aa}. In fact, recent experiments have shown that non-Gaussian noise can indeed be used to tune the performance of a colloidal Stirling engine by shifting the operating speed at which power is maximum \cite{Roy:2021aa}.

The generalisation of our model Eq.~\eqref{model} to higher dimensions includes widely used active particle models such as run-and-tumble particles~\cite{Schnitzer93,TaiCat08}. Here in two dimensions one would have position coordinates $(q_x,q_y)$ and the orientation angle $\theta$ of the active force that receives non-Gaussian noise kicks during tumbling events. With our approach one could, in particular, study the regime where tumbling and diffusion are of comparable strength, rather than the simpler situation where tumbling is so fast that the active force direction becomes effectively slaved to the particle position~\cite{WoiZhaKafLecTai19}. Our method also allows a systematic investigation of non-Gaussian noise effects on activation processes observed in other models for active particle motion~\cite{WexGovRasBel20} and opens up many further fascinating questions, e.g., how non-Gaussian noise affects the selection of the transition states that are traversed during the escape from a metastable state.

\begin{acknowledgements}

We gratefully acknowledge helpful discussions with R.~Jack and K.~Kanazawa. We thank T.~Sano for sharing the simulation data and code of Ref.~\cite{Kanazawa:2020aa}.

\end{acknowledgements}

\begin{appendix}

\section{Dimensionless equation of motion}
\label{SI:dimless}

We consider the overdamped motion of the position coordinate $q$ under the effect of the potential $V(q)$ in one dimension
\be\label{eqmorig}
\gamma\, \dot{q}(t)=-V'(q)+\xi(t),
\ee
where $\gamma$ denotes the friction coefficient and $\xi$ noise from the environment. We assume that $V(q)$ can be expressed as $V(q)=V_0\tilde{V}(q/q_0)$, where $V_0$ and $q_0$ set the energy and spatial scales, respectively, and $\tilde{V}$ is dimensionless. The scale of time can then be set by $t_0=q_0^2\gamma/V_0$. Introducing dimensionless time and position as $\tilde{q}=q/q_0$ and $\tilde{t}=t/t_0$ yields
\be
\frac{\D\tilde{q}(\tilde{t})}{\D \tilde{t}}=-\tilde{V}'(\tilde{q})+\frac{t_0}{\gamma q_0}\xi(t_0\tilde{t}).
\ee
Setting the noise in dimensionless units as
\be
\label{noisedimless}
\tilde{\xi}(\tilde{t})=\frac{t_0}{\gamma q_0}\xi(t_0\tilde{t})
\ee
leads to Eq.~\eqref{model} in the main text, with the tildes dropped from variable names for clarity.

It is straightforward to check that Eq.~\eqref{noisedimless} correctly transforms the specific noise parameters into dimensionless quantities. Assuming first $\xi(t)=\xi_{\rm G}(t)$ as Gaussian white noise with noise intensity $D_0$, i.e.\ $\left<\xi(t)\xi(t')\right>=D_0\gamma^2\delta(t-t')$,
Eq.~\eqref{eqmorig} with $V(q)=0$ implies that $\left<q^2(t)\right>= D_0\,t$ and thus $D_0$ has dimensions $[D_0]=[q_0]^2/[t_0]^2$ as expected for a diffusion coefficient.
The dimensionless noise intensity is then $\tilde{D}_0=D_0 t_0/q_0^2$ and the dimensionless noise has variance $\langle \tilde\xi(\tilde t)\tilde\xi(\tilde t')\rangle = [t_0/(\gamma q_0)]^2 D_0\gamma^2 \delta(t_0(\tilde t - \tilde t')) = (D_0 t_0/q_0^2)\delta(\tilde t - \tilde t') = \tilde D_0 \delta(\tilde t - \tilde t')$.
In the literature our $D_0$ is often written as $2 D_0$ and $\tilde D_0$ as $2\tilde D_0$; we omit the factor of 2 in order to have $\tilde D_0$ directly related to the noise variance.

Secondly, let us assume that the noise $\xi(t)=\xi_{\rm NG}(t)$ is given by the Poissonian shot noise of Eq.~\eqref{psn}. From the fact that $\xi/\gamma$ has the same dimension as $\dot q$, one sees that $[A_j]=[\gamma][q_0]$. The dimensionless amplitudes are thus given by $\tilde{A}_j=A_j/(\gamma q_0)$ consistent with Eq.~\eqref{noisedimless}. In addition, the dimensionless rate is $\tilde{\lambda}_0=\lambda_0t_0$, which preserves $\tilde{\lambda}_0\tilde{t}=\lambda_0 t$; the average number of noise kicks is therefore unaffected by the change to dimensionless units as it must be.

\section{Large deviation form of the path probability for non-Gaussian noise}
\label{SI:pi}

\subsection{The cumulant generator for non-Gaussian noise with independent stationary increments}

For the Poissonian shot noise of Eq.~\eqref{psn}, we see that the increments $\xibar{s} \equiv \int_s^{s+\Delta t} \xi_{\rm NG}(s)\D s$ over a small time step $\dt$ are all independent and assume the values $\xibar{s}=A$ with probability $\lambda_0 \dt$ and $\xibar{s}=0$ otherwise. The characteristic function of a given increment is thus
\be
\label{si:cf1}
\langle e^{ig(s)\xibar{s}}\rangle&=&\langle e^{ig(s)A}\lambda_0\dt+1(1-\lambda_0\dt)\rangle\nonumber\\
&\approx&  \exp\left(\lambda_0\dt\langle e^{ig(s)A}-1\rangle\right),
\ee
with the remaining average taken over the amplitude distribution $\rho_0$. In general one wants the noise to have zero mean. Subtracting the constant average $\langle \xi(t)\rangle$ to enforce this in Eq.~\eqref{psn} gives an extra term $-ig(s)A$ inside the average of Eq.~\eqref{si:cf1}. Adding also in Eq.~\eqref{psn} a Gaussian noise contribution with variance $D_0$ we obtain
\be
\label{si:cf2}
\langle e^{ig(s)\xibar{s}}\rangle=\exp\left(-\frac{D_0}{2}g(s)^2\dt+\lambda_0\phi_0(ig(s))\dt\right),\nonumber\\
\ee
where $\phi_0$ is given as
\be
\phi_0(k)=\int\D A\,\rho_0(A)\left( e^{k A}-k A-1\right).
\ee
As explained in the main text we will find it useful to write $\rho_0(A)=\rho(A/a_0)/a_0$ in terms of a characteristic scale $a_0$ and a base distribution $\rho$, normalized so that $\int\D x\, x^2\rho(x)=1$. For $\phi_0$ this scaling implies $\phi_0(k)=\phi(k a_0)$, where
\be
\phi(u)=\int\D A\,\rho(A)\left( e^{u A}-u A-1\right).
\ee
and the normalization of $\rho$ simplifies the non-Gaussian noise variance to $\lambda_0 \langle A^2\rangle=\lambda_0 a_0^2$. Considering a whole noise trajectory and the continuum limit $\dt\to 0$ recovers the noise cumulant generator Eq.~\eqref{cflevy}.

\subsection{MSR action functional}

In order to develop a path integral description of the dynamics Eq.~\eqref{model} we again consider first a discretization into small time steps $\dt$. Using an Ito convention Eq.~\eqref{model} can be discretized as
\be
\label{si:discret}
q(s+\dt)=q(s)+\dt\,V'(q(s))+\xibar{s}.
\ee
Enforcing the dynamics Eq.~\eqref{si:discret} at every time step with delta functions, we can express the probability of a path $[q]=(q(0),q(\dt),\ldots,q(t))$ with fixed $q(0)$ as a product 
\begin{equation}
P[q]=\left\langle \prod_{s=0}^{t-\dt} \delta(q(s+\dt)-q(s)+\dt\,V'(q(s))-\xibar{s})\right\rangle.
\end{equation}
The average is over the noises $\xibar{s} \equiv \int_s^{s+\dt} \xi(s)$ and can be done independently for each time step. Fourier transforming one such step gives
\begin{widetext}
\be 
\int\frac{\D \uu(s)}{2\pi} e^{-i\uu(s)
[q(s+\dt)-q(s)+\dt\,V'(q(s))]}\langle e^{i\uu(s)\xibar{s}} \rangle=\int\frac{\D \uu(s)}{2\pi} e^{-i\uu(s)[q(s+\dt)-q(s)+\dt\,V'(q(s))]-\frac{D_0}{2}g(s)^2\dt+\lambda_0\phi(ig(s)a_0)\dt}\nonumber\\
\ee
using Eq.~\eqref{si:cf2}. Collecting the contributions from all time steps and taking $\dt\to 0$ gives the path probability in terms of a Martin-Siggia-Rose (MSR)-type action $S[q,\uu]$ \cite{Martin:1973aa,Hertz:2016aa}:
\be 
P[q]&=&\int\!\mathcal{D}\! \left[\frac{\uu}{2\pi}\right]e^{-S[q,\uu]}
\\
S[q,\uu]&=&\int_0^t \D s\,\left\{i\uu(s)[\dot q(s)+V'(q(s))]+\frac{D_0}{2}g(s)^2-\lambda_0\phi(ig(s)a_0)\right\}
\ee

\end{widetext}

\subsection{Rescaling the noise parameters}
\label{App_Sec_rescaling}

\begin{figure}
\begin{center}
\includegraphics[height=5cm]{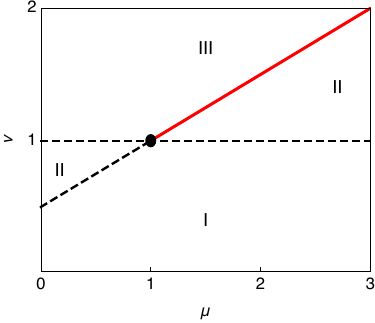}
\caption{\label{Fig:regimes} Different noise regimes arising under the scaling $\lambda_0=\lambda/\epsilon^\mu$, $a_0=a\,\epsilon^\nu$ in the limit $\epsilon\to 0$}
\end{center}
\end{figure}

The seminal Kramers escape rate for Gaussian noise ($\lambda=0$) can formally be derived from the theory of large deviations that is applicable in the weak-noise limit $D_0\to 0$. Fluctuations around the most likely path from one metastable state to another are then small and the typical path can be obtained by making the action $S[q]$ stationary w.r.t.\ $q(s)$ and $\uu(s)$. In order to analyse such a weak-noise regime, we introduce a dimensionless scaling parameter $\epsilon$ and rescale $D_0$ as
\be
D_0=D\,\epsilon,
\ee
such that the weak-noise regime is equivalent to taking $\epsilon\to 0$. Setting $\delta S/\delta \uu(s)=0$ (still for $\lambda=0$) gives $D\epsilon g=i(\dot q+V')$, showing that in the low-$\epsilon$ limit one needs to scale  $g=\tilde{g}/\epsilon$. The action then becomes
\be
\label{si:action}
S[q,\tilde{\uu}]=\frac{1}{\epsilon}\int_0^t \D s\,\left\{i\tilde{g}[\dot q+V'(q)]+\frac{\tilde{g}^2}{2}-\lambda_0 \epsilon\phi(i\tilde\uu a_0/\epsilon)\right\}\nonumber\\
\ee
Without the non-Gaussian term this already has the desired scaling with $\epsilon^{-1}$ that shows how path fluctuations away from the most likely path become exponentially suppressed for small $\epsilon$.

For nonzero $\lambda$ the task now is to identify a scaling regime that achieves the same result for the non-Gaussian contribution. The non-Gaussian term $\lambda_0 \epsilon\phi(i\tilde{g}a_0/\epsilon)$ in Eq.~\eqref{si:action} suggests the scaling $\lambda_0=\lambda/\epsilon$, $a_0=a\, \epsilon$ considered in the main text (cf. Eq.~\eqref{rescaling2}). We show now that this is in fact the only scaling that preserves all non-Gaussian noise features, by considering general scaling exponents
\be
\lambda_0=\lambda/\epsilon^\mu, \qquad\qquad a_0=a\,\epsilon^\nu.
\ee
Expanding then the function $\phi$ yields
\be
\label{si:phiscal}
\phi(i\tilde{g}/\epsilon) = \frac{a_0^2\langle x^2\rangle}{\epsilon^2}\frac{(i\tilde{g})^2}{2!}+\frac{a_0^3 \langle x^3\rangle}{\epsilon^3}\frac{(i\tilde{g})^3}{3!}+ \ldots
\ee
so that the $O(\tilde{g}^n)$ term of $\lambda_0 \epsilon \phi$ scales as $\epsilon^{1-\mu+n(\nu-1)}$. The exponents $\mu$, $\nu$ thus define different scaling regimes for $\epsilon\to 0$ as shown in Fig.~\ref{Fig:regimes}.

In regime I, all orders ($n\geq 2$) in $\tilde g$ diverge as $\epsilon\to 0$. In regime II, there are always some higher orders that diverge as $\epsilon\to 0$, while in regime III all orders scale to zero as $\epsilon\to 0$ so that one effectively recovers the case $\lambda_0=0$. For the particular combination $\nu=\frac{1}{2}(\mu+1)$ with $\mu>1$ (red line in Fig.~\ref{Fig:regimes}) only the $\tilde{g}^2$ term remains in Eq.~\eqref{si:phiscal} as $\epsilon\to 0$. The non-Gaussian noise strength  $\lambda_0 a_0^2 \propto \epsilon \to 0$ here, so this is a valid weak noise-limit but one that reduces to effective Gaussian noise. Only for $\mu=\nu=1$ do all orders in $\tilde g$ remain in Eq.~\eqref{si:phiscal} as $\epsilon\to 0$. This is therefore the scaling we adopt: it represents a genuine weak-noise limit of our generic noise, since the noise variance is $D_0+\lambda_0 a_0^2=(D+\lambda a^2)\epsilon\propto \epsilon$ while the infinite hierarchy of noise cumulants is retained. The action then simplifies to $S[q,\tilde{g}]=\tilde{S}[q,\tilde{g}]/\epsilon$ with
\be 
\label{si:tildeS}
\tilde{S}[q,\tilde{\uu}]=\int_0^t \D s\,\left\{i\tilde{g}[\dot q+V'(q)]+\frac{\tilde{g}^2}{2}- \lambda \phi(i \tilde{g}a)\right\}\nonumber\\
\ee
and contains $\epsilon$ only through the overall scale $\epsilon^{-1}$ as desired. The path probabilities are as before except for the scaling of the conjugate variables, 
\be
\label{si:pathprob2} 
P[q]=\int\mathcal{D}\left[\frac{\tilde{g}}{2\pi \epsilon}\right] e^{-\tilde{S}[q,\tilde{g}]/\epsilon}.
\ee

\subsection{Saddle-point integration}

With the above large deviation form of the path probability, a path-integral expression for the propagator of the dynamics Eq.~\eqref{model}, i.e., the probability of reaching a given $q(t)$ from some $q(0)$, can be obtained by integrating over all paths with those end points. For $\epsilon\to 0$, this propagator is dominated by the path that makes the action Eq.~\eqref{si:tildeS} stationary, which can be found by solving the associated Euler-Lagrange equations for $q(s),\tilde{g}(s)$. However, these presume continuous paths and we find that for some non-Gaussian noise types such solutions do not exist for low $\lambda$. But we can obtain a description that extends to this more difficult regime by first eliminating $\tilde{g}$ in Eqs.~\eqtoref{si:tildeS}{si:pathprob2} by saddle point integration in the weak noise limit $\epsilon\to 0$. Technically we discretize into small time intervals $\dt$ and take $\epsilon\to 0$ first, then $\dt\to 0$. The stationarity condition
\be
0=i[\dot q+V'(q)]+\tilde{g}-i\lambda a\phi'(i\tilde{g}a)
\ee
shows that $\tilde{g}$ is imaginary at the saddle point, so in terms of $k=i\tilde{g}$ the resulting contribution to the action can be written as 
\be
\label{si:pi_def}
\lagr(f) &=& \max_k\{k f-k^2/2-\lambda\phi(a k)\}\nonumber\\
&=& \max_k\{k f-\psi(\mom)\}
\ee
with $f=\dot q+V'(q)$. The maximum rather than minimum appears here because of the saddle structure of the stationary point. One can check that this result remains valid even when $\phi$ has singularities on the real axis; the maximum in Eq.~\eqref{si:pi_def} then has to be taken over the range where $\phi$ remains non-singular. In our examples in Table~\ref{Fig_phis} such singularities occur for the Gamma and exponential noise amplitude distributions. In contrast, distributions with tails decaying faster than exponentially do not exhibit such singularities; see the constant modulus example in Table~\ref{Fig_phis}.

\section{Analysis of the escape behaviour}

\label{SI:escape}

\subsection{Parameter regime for non-Gaussian effects}

To understand the reduction in $\Smin$ as a function of $\lambda$ and $a$ we write Eq.~\eqref{k_def} with Eq.~\eqref{psi} as
\be
\kn^*=\atilde \mom^*,\qquad V'(q)=\frac{1}{2\atilde}\,\kn^*+\lambdatilde \atilde\frac{\pphi_0(\kn^*)}{\kn^*}.
\ee
The terms on the right are both positive so if either of the prefactors are large ($1/\atilde\gg 1$ or $\lambdatilde \atilde\gg 1$) this will force $\kn^*$ to be small. Now for small arguments $\phi(u)\approx u^{2}/2$ and one obtains $\kn^*=2V'/(\atilde^{-1}+\lambdatilde \atilde)$. Bearing in mind that $\mom^*=\kn^*/\atilde$, the minimum action $\Smin$ from Eq.~\eqref{Sgen} then takes the Gaussian value, $\Smin \approx \Sg$. Likewise, the instanton in this regime will assume the Gaussian shape, since $\psi(\mom)\approx (1+\lambdatilde \atilde^2)\mom^2/2$ and Eq.~\eqref{instanton} yields $\dot{q}=V'(q)$. 

Summarizing, we predict Gaussian behaviour when $1/\atilde\gg 1$ or $\lambdatilde \atilde\gg 1$. Conversely, to see non-Gaussian noise effects we need the noise amplitude to lie in the range $1\ll \atilde \ll 1/\lambdatilde$; such a range exists for $\lambdatilde \ll 1$. These predictions are consistent with the data shown in Fig.~\ref{Fig_action}a,b.

\subsection{Comparison with literature results for the action in special cases}

We briefly review literature results where analytical predictions for the effective action $S_{\rm min}$ of the escape problem have been obtained for special cases of our general non-Gaussian noise as defined in Eqs.~\eqqref{cflevy}{phi}.

In \cite{Van-den-Broeck:1984aa,Sancho:1985aa,Porra:1993aa}, one-sided Poissonian shot noise with exponentially distributed amplitudes was considered, which corresponds to $\rho(x)=e^{-x}/2$ for $x>0$ once we impose our normalization $\int\D x\, x^2\rho(x)=1$. With Eq.~\eqref{phi} we obtain the associated moment generator
\be
\phi(u)=\frac{u^2}{2(1-u)}
\ee
and we also have $D=0$ due to the absence of a Gaussian component. The condition for $k^*$, Eq.~\eqref{k_def}, is thus
\be
V'(q)=\lambda a^2\frac{k^*}{2(1-ak^*)}
\ee
and solving for $k^*$ yields the action with Eq.~\eqref{Sgen}
\be
\Smin=\int_{q_a}^{q_b} \D q\,\frac{2V'(q)}{\lambda a^2+2aV'(q)},
\ee
which has been obtained in \cite{Van-den-Broeck:1984aa,Sancho:1985aa,Porra:1993aa}.

In \cite{Billings:2008aa}, the authors consider one-sided Poissonian shot noise with constant one-sided amplitudes, where $\rho(x)=\delta(x-1)$ and thus
\be
\phi(u)=e^u-u-1.
\ee
In this case, Eq.~\eqref{k_def} cannot be solved in closed form for $k^*$. Rearranging Eq.~\eqref{k_def} with $D=0$ yields $k^*$ as the solution of
\be
k^*=\frac{1}{a}\ln\left(1+\left(a+\frac{V'(q)}{\lambda}\right)k^*\right),
\ee
and the action obtained via Eq.~\eqref{Sgen} recovers the result in  \cite{Billings:2008aa}.

In \cite{Gera:2021ve}, the authors consider a combination of Gaussian noise and two-sided Poissonian shot noise with exponentially distributed amplitudes, which is one of the cases considered in the main text. Rearranging Eq.~\eqref{k_def} for the type B case of Fig.~\ref{Fig_phis} yields $k^*$ as the solution of
\be
k^*=\frac{2V'(q)}{D+\lambda a^2/(1-a^2k^{*2})}.
\ee
Eq.~\eqref{Sgen} with this expression for $k^*$ matches the result obtained in \cite{Gera:2021ve}, bearing in mind the difference by a factor 2 due to the different noise intensity conventions used.

\end{appendix}

%\bibliography{../fluctuations,../non-Gaussian_noise.bib}

\end{document}